# Deep Learning-based Kinetic Analysis in Paper-based Analytical Cartridges Integrated with Field-effect Transistors

Hyun-June Jang[1,2]‡, Hyou-Arm Joung[3]‡, Artem Goncharov[3], Anastasia Gant Kanegusuku[4], Clarence W. Chan[5], Kiang-Teck Jerry Yeo[5,6], Wen Zhuang[1,2], Aydogan Ozcan[3,7,8,9]*, Junhong Chen[1,2]*

[1]Pritzker School of Molecular Engineering, University of Chicago, Chicago, IL 60637, USA

[2]Chemical Sciences and Engineering Division, Physical Sciences and Engineering Directorate, Argonne National Laboratory, Lemont, IL 60439, USA

[3]Department of Electrical and Computer Engineering, University of California, Los Angeles, CA 90095, USA

[4]Department of Pathology and Laboratory Medicine, Loyola University Medical Center, Maywood, IL 60153

[5]Department of Pathology, The University of Chicago, Chicago, IL 60637, USA

[6]Pritzker School of Medicine, The University of Chicago, Chicago, IL 60637, USA

[7]Department of Bioengineering, University of California, Los Angeles, CA 90095, USA

[8]California NanoSystems Institute (CNSI), University of California, Los Angeles, CA 90095, USA

[9]Department of Surgery, David Geffen School of Medicine, University of California, Los Angeles, CA 90095, USA

‡These authors contributed equally to this work

*Correspondence: ozcan@ucla.edu, junhongchen@uchicago.edu

**Summary**

This study explores the fusion of a field-effect transistor (FET), a paper-based analytical cartridge, and the computational power of deep learning (DL) for quantitative biosensing via kinetic analyses. The FET sensors address the low sensitivity challenge observed in paper analytical devices, enabling electrical measurements with kinetic data. The paper-based cartridge eliminates the need for surface chemistry required in FET sensors, ensuring economical operation (cost < $0.15/test). The DL analysis mitigates chronic challenges of FET biosensors such as sample matrix interference, by leveraging kinetic data from target-specific bioreactions. In our proof-of-concept demonstration, our DL-based analyses showcased a coefficient of variation of < 6.46% and a decent concentration measurement correlation with an $r^2$ value of > 0.976 for cholesterol testing when blindly compared to results obtained from a CLIA-certified clinical laboratory. These integrated technologies can create a new generation of FET-based biosensors, potentially transforming point-of-care diagnostics and at-home testing through enhanced accessibility, ease-of-use, and accuracy.

**INTRODUCTION**

The landscape of scientific research and technological innovation has witnessed an extraordinary convergence of diverse disciplines, fostering profound advancements across an extensive spectrum of domains. In this dynamic milieu, the emergence of machine learning methodologies heralds a transformative epoch that fundamentally reshapes the contours of medical diagnostics. What is particularly remarkable is the far-reaching impact of this paradigm shift, transcending customary boundaries of laboratory environments to encompass at-home testing, point-of-care (POC) diagnostics, and a broad array of real-world applications. Advanced machine learning methods such as neural networks have been recently emerging in clinical diagnostics with applications in histology, biosensing technologies, and serodiagnosis of cardiovascular diseases, among others.[1-4] Neural networks can learn from highly multiplexed and non-linear responses of POC sensors and accurately quantify analyte concentrations despite cartridge-to-cartridge variations and the noise present in biological samples (e.g., matrix effect).[5,6]

Traditional biomedical disciplines have extensively relied on the foundational principles of optical detection techniques and bioimaging, capturing the light signals associated with targeted biomarkers.[7-9] Real-time bio-signal measurements and bioimaging technologies offer distinct advantages, including internal quality control through data-driven image analysis[1], improved precision and accuracy[2, 4-5], and expedited detection capabilities.[9] Despite their elegance, optical methodologies are partially encumbered by inherent limitations, particularly in the context of POC and at-home testing scenarios. Some of these limitations include high reagent costs and the need for trained experts/users, instigating an exploration of innovative alternatives in diagnostic platforms.[10]

Field-effect transistors (FETs), originally designed for electronic circuitry, have undergone a remarkable evolution, transforming from conventional electronic components into highly sensitive transducers capable of real-time and label-free detection of a diverse array of analytes with unparalleled sensitivity.[11] Despite decades of research, FET-based biosensors remain in the proof-of-concept stage, lacking successful market products.[12,13] The realization of their commercial potential, however, has been hindered by a myriad of challenges: batch variation (i.e., reproducibility)[14], sample matrix effects[10], and packing requirements associated with wet environments on the sensing surface[15], as well as susceptibility to risks such as leakage and contamination of the sensing zone[16]. Recently, a set of FET-based sensor designs has exhibited promising results in overcoming matrix effects and batch variations in sensing metal ions[14,17,18] by leveraging machine learning. However, these approaches have not been extended to sensing within complex testing environments, such as human blood or plasma, which includes divergent patient-specific proteins, cellular compositions, and ion variations.

While the conventional real-time FET detection captures a snapshot of a bioreaction via monitoring continuous changes in drain current ($I_D$) at a fixed gate voltage ($V_G$) application or threshold voltage ($V_{th}$), its advantages in terms of accuracy and precision have not been thoroughly elucidated as the gate-dependence of the kinetic data is not captured. Additionally, these FET biosensors mandate the incorporation of additional reagent control mechanisms to address issues related to the Debye length[19,20] for physiological media (e.g., capture-release method[21] associated with reactions, washing, and measurements under controlled buffers), often facilitated through microfluidic systems for better usability. However, the fabrication of such microfluidic channels frequently encounters challenges, including time-consuming and labor-intensive manufacturing processes, diminished yield and the intricate nature of configuring microfluidic setups involving pumps and valves.[22-24] These constraints can complicate operations, contradicting the convenient and straightforward use of POC diagnostic tools, such as lateral-flow assays (LFAs).[24-25]

In this manuscript, we embark on an exploration of the synergism between FET biosensors, paper-based analytical devices, and the computational power of deep learning (DL). We first demonstrate the integration of a porous sensing membrane (PSM), a key component of LFAs, with FET sensors. The platform features a one-step operation for sample injection, a disposable cartridge for introducing plasma samples (cost < $0.15 per test), and reusable FET sensors. The PSM incorporates dried sensing components that produce electroactive enzymatic signals, such as protons specific to the target biomarkers in plasma. The FET sensor characterizes all-encompassing, target-specific kinetic data occurring within the cartridge through FET transfer curves. Finally, the DL-based analysis of the obtained kinetic data effectively addresses issues related to sample matrix effects and varying rates of chemical reactions from test-to-test, and accurately quantifies the target analyte concentration from the captured data. We showcased a proof-of-concept operation of this platform for cholesterol testing with patient plasma. Cholesterol concentrations blindly predicted by DL exhibited a correlation

($r^2$) of > 0.976, with a coefficient of variation (CV) of < 6.42%, when blindly compared against results obtained from a CLIA-certified clinical laboratory for the same samples.

## RESULTS

**Operating Principle.** Our innovative diagnostic platform employs a new dry chemistry approach for FET biosensors. The paper analytical cartridge, housing the desiccated sensing components in the PSM, is electrically connected to the FET (Figure 1a). This synergetic integration offers several advantages: 1) imposing higher sensitivity to paper analytical devices empowered by electrical measurements, 2) enabling the measurement of kinetic data, 3) eliminating the complexities associated with laborious wet chemistry procedures for functionalizing biomolecules on the FET sensing surface, 4) removing the need for traditional microfluidic systems to regulate reagents, 5) facilitating cost-effective testing (~$0.15 per cartridge, Table S1), 6) mitigating sample matrix effects, 7) achieving easy miniaturization, and 8) ensuring a prolonged shelf-life (the shelf-life of LFAs is up to 2 years[25]).

As a proof-of-concept demonstration for our diagnostic approach, we selected cholesterol, a standard biomarker in annual blood testing at clinics. The PSM was desiccated with enzymes such as cholesterol esterase (COE), cholesterol oxidase (COx), and peroxidase (POx), along with surfactants, stabilizers, and buffers, eliminating the need for additional functionalization steps (Figure 1b). An ion-sensitive sensing electrode (SE), such as indium-tin-oxide (ITO), was positioned beneath the PSM within the cartridge, with ~50 μm of physical spacing. Injecting 20 μL of plasma into the cartridge's inlet established a connection between the ITO and the FET gate (Figure 1a). Once the plasma contacted the PSM, surfactants broke down lipoproteins, and a series of enzymatic reactions produced protons released into the physical spaces between the PSM and ITO. The real-time release of protons resulting from a series of enzymatic reactions specific to the cholesterol concentration in the plasma was continuously recorded in FET transfer curves repeatedly measured over 5 minutes (Figure 1c). These transfer curves were transformed into a 2D heatmap, encapsulating all enzymatic kinetic details characterized in the sum of transfer curves. The DL analysis further optimized the subset of kinetic signals carrying concentration-specific data to quantify cholesterol concentrations in patient plasma samples (Figure 1c).

**Proton Specificity.** Proton was the target signal of interest in response to cholesterol. The intrinsic ITO exhibited a Nernstian response of 52.8 mV/pH with an $r^2$ of 0.997 and CV < 1.3% (Figure S1a) without any changes in the transconductance ($G_m$) value over all pH ranges (Figure S1b). Due to the significantly higher input impedance of the FET compared with that of an ITO remote gate module[26-28], there were no changes in $V_{th}$ with increasing contact areas as a result of increasing media volume size (from 20 to 100 μL) on the ITO surface (Figure S1c). An insignificant drift of 8 μV/min, measured over 30 minutes (Figure S1d), suggested that the ITO sensing electrode was highly stable for translating enzymatic reactions specifically.

**FET Data.** Upon injecting human plasma into the cartridge, distinct real-time signal patterns were observed in response to varying cholesterol levels, as indicated by changes in the $V_{th}$ ($\Delta V_{th}$) relative to the $V_{th}$ of lipoprotein-free cholesterol plasma (Figure 2a). The protons generated by each electroactive enzymatic reaction decreased $V_{th}$ levels of n-type FETs due to the positive surface potentials applied on the ITO from protons. The initial $V_{th}$ of the cartridges was largely influenced by the pH and ion concentration of human plasma, as well as batch variations in cartridges, along with diverse proteins and components in plasma that could cause non-specific binding on the ITO surface. Despite injections of different plasma samples, the initial $V_{th}$ values tended to overlap (Figure 2a). This could be attributed to our cartridge design, which incorporated a ~50 μm air gap between the PSM and the ITO electrode (inset of Figure 2a). The presence of this air gap played a pivotal role in facilitating the mixing process between the PSM and plasma samples. It ensured that the mixing process occurred before the original plasma came into direct contact with the ITO electrode (inset of Figure 2a). As a result, any sample matrix effects were significantly diluted by the potent buffer components that were already desiccated within the PSM. Consequently, the PSM efficiently transmitted purified electroenzymatic signals to the ITO surface.

Figure 2b further supports the fluidic dynamics in the cartridge described above by demonstrating the controlled initial $V_{th}$ of each cartridge after the injection of different random human plasma samples using various buffer solutions, such as phosphate-buffered saline (PBS), piperazine-N,N′-bis(2-ethanesulfonic acid) (PIPES), and PIPES/PBS (5:5 ratio), dried onto identical PSMs, with the remaining PSM components being consistent. Small variations in the initial $V_{th}$ levels are shown for each group of cartridges dried with PBS, PIPES/PBS, and PIPES (CV < 10%). Even lipoprotein-free plasma samples with a pH of 5 tend to be controlled by the buffer solution in the PSM of the cartridge (Figure 2b). For subsequent experiments, we selected PBS (pH 7.4) as the buffer component for the PSM to minimize the pH disparity between plasma (pH 7.35 to 7.45) and the dried buffers.

We further optimized the concentrations of COE, COx, and POx for the PSM by evaluating correlations between the enzyme concentrations and the cholesterol signals (Figure S2). Concentrations of enzymes exceeding 200 U/mL displayed saturated correlations, leading us to choose 300 U/mL enzyme concentrations for COE, COx, and POx in all subsequent experiments. However, without the POx enzyme, no correlative sensing signals were obtained, implying that POx played a critical role in producing electroactive enzymatic signals such as protons (Figure S3). It is noted that the enzyme solutions lacked long-term shelf life without drying them on the PSM, as shown by a large drift in the initial $V_{th}$ levels of the enzyme solution over time (Figure S4).

We further discovered that the use of bovine serum albumin (BSA) coating on the ITO electrode significantly mitigated sample matrix effects without compromising the detection signals (Figure 2c). In contrast, bare ITO electrodes without BSA coating exhibited larger shifts in electroenzymatic signals, even for free cholesterol plasma samples. This might be due to the remaining proteins and ions in lipoprotein-free plasma causing non-specific signals by interacting with the bare ITO surface.

The LOD achieved by our detection platform was assessed in Figure 2d using diluted clinical plasma with lipoprotein-free plasma, ensuring controlled conditions. The estimated LOD for cholesterol is determined to be 28.5 µg/dL (737 nM). This LOD range is exclusively demonstrated by conventional electrochemical detection methods employing sophisticated and complex device fabrications, including the utilization of nanomaterials and mediators.[29] The highly sensitive attribute of FET sensors extends this remarkable LOD range to paper-based analytical devices, highlighting the efficacy of the platform even without the need for functionalization.

Lastly, in Figure 2e, we present the distribution of $\Delta V_{th}$ values for cholesterol concentrations measured from patient plasma samples across 178 cartridges produced during 15 different sub-batch fabrications. Notably, when enzymes were absent from the PSM, no correlated detection signals were observed, leaving sample matrix effects the only reason for minor variations in $\Delta V_{th}$ (Figure S5). While there was a degree of correlation between cholesterol levels and $\Delta V_{th}$ values, relying solely on $\Delta V_{th}$ (which is conventionally used in FET sensor analysis to determine target-signals) as the sensor readout presented major challenges. When applied to our dataset, this standard approach displayed a low $r^2$ of 0.808 (Figure 2e) and a large CV of up to 22.8% within a clinically relevant range (i.e., 100-150 mg/dL, Figure 2f), confirming the challenges of FET biosensor performance on physiological samples, despite optimizations of the diverse fabrication factors listed above.

The conventional FET analysis, focusing solely on $\Delta V_{th}$ (Figure 2e) by considering the endpoint and initial point, falls short of providing a comprehensive understanding of variations during enzyme reactions influenced by the time-dependent enzyme reaction rate (Figure 3a) and sample matrix effects. Moreover, the real-time FET measurement (Figure 2a), capturing a cumulative representation of specific snapshots at particular moments over time, also offers limited information about enzyme reactions. For instance, the $G_m$ values, which cannot be obtained from real-time measurement approach in Figure 2a, exhibit significant variations during reactions (Figure 3b). These variations are influenced by factors such as the rate of enzyme reactions, the mixing process within the cartridge, and the presence of sample matrix effects. Interpreting these dynamic behaviors for each specific case can be challenging, underscoring the need for more advanced data-driven analytical techniques, such as DL, to comprehensively capture and interpret the dynamic nature of these biochemical processes, as detailed in the subsequent sections.

**Design of DL-based Signal Analysis.** DL benefits from the universal function approximation power of neural networks to harness the complex non-linear kinetic data from the FET sensor to measure the analyte concentrations. Here, we employed DL-based analysis and neural networks for two key objectives: 1) optimizing the subset of kinetic signals carrying concentration-specific information, and 2) quantifying the target analyte concentrations in patient plasma samples. The DL models were structured as fully-connected shallow networks with three hidden layers, utilizing continuously measured FET transfer curves as input data. Both the kinetic data input and the network architecture underwent optimization through a 4-fold cross-validation on the validation set of plasma samples (refer to the Data Processing and Deep Learning Analysis section for detailed procedures). The optimized network was subsequently tested on 30 additional samples from the testing set, never used before.

For DL analysis, transfer curves measured over 5 minutes for each plasma sample were transformed into a 2D heatmap (Figure S6). This heatmap for each test visually represented all enzymatic kinetic details, encompassing characteristics observed in the sum of raw transfer curves (Figure S6a), such as potential drifts during the measurement, initial $V_{th}$, and changes in $V_{th}$, electronic mobility, and $G_m$ due to enzyme reactions (Figure S6b). Notably, our preliminary observations underscored the significance of subtracting the initial transfer curve data from the raw heatmap in Figure S6b. The initial transfer curves typically serve as a baseline for conventional FET analysis to measure the relative change in target-specific signals, which could be significantly affected by pH, ion concentrations, and sample matrix effects. Thus, the raw heatmap (Figure S6b), after subtracting the initial transfer curve properties and referred to as the signal heatmap (Figure 3c), encapsulated pure kinetic information primarily associated with enzymatic reactions, and remained, by and large, unaffected by interference from varying pH levels in plasma samples. Using this signal heatmap as input resulted in a substantial improvement in the neural network inference, reducing the CV from 20.1% to 8.5% and increasing the $r^2$ from 0.698 to 0.904 (see Figure 3d).

DL was further applied to optimize the subset of kinetic signals containing concentration-specific information within the signal heatmap. This optimization considered both the size of the $V_G$ window (i.e., within the 0-3 V range, Figure 4a-c) and the time window (i.e., within the 14-343 s range, Figure 4d-f). The selection of the optimal model was based on achieving the lowest mean square error (MSE) and the highest $r^2$ values when comparing predicted and ground truth cholesterol concentrations for samples from the validation dataset, with variations in the sizes of $V_G$ windows. Consequently, the optimal $V_G$ window was identified to be between 1.15 V and 2.45 V (Figure 4b). The predicted concentrations within this optimal $V_G$ subset still exhibited a high CV of 20.7% and a relatively low $r^2$ of 0.907. With a fixed $V_G$ window, an optimal time subset was determined to be within the range of 91-119 s (Figure 4e). The network utilizing the optimized subset in Figure 4e demonstrated improved quantification performance on the same validation set in Figure 4c, achieving an $r^2$ of 0.954 and a CV of 11.4% (Figure 4f). This final optimized network was further utilized to generate blind testing results using plasma samples never seen before, which will be detailed in the next section.

**Blinded Testing Results.** The optimized neural network model, incorporating the optimal architecture and refined kinetic data input (refer to the Data Processing and Deep Learning Analysis section of the Methods for detailed model architecture), underwent blind testing to quantify cholesterol concentrations across 30 clinical plasma samples from three distinct testing batches (Figure 4g). For each of the three testing batches, we trained optimized models separately using samples from the same batch to minimize inter-batch variability. The blind testing predictions exhibited a high correlation with the ground truth cholesterol values, with an $r^2$ value exceeding 0.976 for all three batches (Figure 4h). Additionally, the neural network models' inference results demonstrated low variations in predictions, with a maximum CV of 6.46% over different cholesterol concentration ranges (Figure 4i). Importantly, blind testing predictions from the models using the optimal subset in the signal heatmap (Figure 4g) outperformed other models, including the model that used the entire raw heatmap (Figure S7a), the model with a 2x larger window size than the optimal subset in the signal heatmap (Figure S7b), and the model using the optimal subset in the raw heatmap (Figure S7c). Therefore, this optimized concentration inference method not only better utilized the enzymatic reaction kinetics of our FET-based sensor but also improved the robustness of the network predictions, making them more resistant to variations induced by sample matrix effects.

For the same blind testing set, a single model trained on samples from all three baches (Figure S8) had inferior accuracy ($r^2$: 0.886 excluding outliers) and precision (CV of 10.85 %). Higher variations of this single neural network model between different batches originate from additional variabilities in PSM and reagent batches used during different testing days. In future iterations of assay development, incorporating batch-specific information along with the sensor data can be used to create a more robust inference model that generalizes to different batches with the same superior performance.

**Scalability to Immunoassay.** Our detection platform has the potential to be adapted for immunoassays, which holds significant promise in a wide range of biomedical applications. To illustrate this adaptability, we conducted a proof-of-concept experiment involving electroactive enzymatic signaling on our platform (Figure S9). In this experiment, we utilized the interaction between horseradish peroxidase (HRP)-labeled anti-mouse IgG (Ab-HRP), hydrogen peroxide ($H_2O_2$), and 3,3',5,5'-tetramethylbenzidine (TMB) — a combination that has long been established and widely used in conventional enzyme-linked immunosorbent assays (ELISA) for detecting biomarkers in sandwich immunoassays using the resultant colored product of TMB.[30] TMB and HRP reactions also produce protons[31], which serve as a target signal in our detection platform. This phenomenon is depicted in Figure S9, where increased concentrations of Ab-HRP decrease $V_{th}$ levels of the FET upon the injection of $H_2O_2$ (Figure S9a) or TMB (Figure S9b). TMB and HRP signaling can be integrated into the LFA framework, complemented by zones dedicated to capture antibody, detection antibody, and chemical substrate. This design could be particularly beneficial for immunoassays that require high sensitivity, such as troponin I and metabolite assays, as well as for the rapid detection of infectious diseases.

**DISCUSSION**

Our integration of an LFA component, specifically the PSM, with a FET yields synergistic benefits and multiple advantages, effectively overcoming limitations inherent in each of the two components when used individually as diagnostic platforms. The FET sensing mechanism can potentially be used to enhance the sensitivity of paper-based analytical devices and enable the measurement of kinetic data within the PSM. On the flip side, the intricate surface chemistry required for FET sensors to immobilize biomolecules like enzymes and antibodies is replaced by the straightforward dry chemistry of the LFA technology. Furthermore, building upon the utilization of an electroactive enzymatic signal, such as protons, as opposed to label-free FET detection, our approach eliminates the requirement for conventional microfluidic systems to regulate reagent supply to FET sensors, effectively addressing Debye length issues. This strategic choice enables a simplified operation down to a single sample injection step, eradicating the need for a complex microfluidic system. This integration goes a step further in mitigating critical commercialization risks associated with FET sensors, including shelf-life, production costs, and contamination of the FET sensing surface, all of which have hindered the widespread adoption of FET biosensors in the market.

Despite effectively addressing the aforementioned risks, some challenges still persist, particularly in terms of reliability, impaired by sample matrix effects and batch-to-batch variations. The enhanced sensitivity of FET sensors, advantageous for detecting target signals, introduces susceptibility to interference from non-specific binding or unintended interactions, which can complicate the interpretation of results and impact the accuracy of the sensor. Various strategies, such as surface modifications, advanced coatings, or the use of blocking agents, have been explored to enhance the specificity of FET sensors. However, the translation of their performance from laboratory-scale studies to practical applications remains a challenge.

Our DL-based analysis, coupled with a meticulous optimization process, proves instrumental in mitigating variation issues arising from sample matrix effects and reaction rates. This approach enhances the ability to discern and interpret complex interactions of the testing environment within the DL model, allowing for a more precise analysis of the intricate but insightful kinetic FET data. While our DL techniques showed competitive cholesterol quantification in samples within the same batch (i.e., models for each of the three batches were trained independently), the scalability of our DL-based FET sensor between different batches was limited (i.e., a single model trained on all three batches showed inferior performance, $r^2 < 0.9$) due to additional factors affecting inter-batch repeatability. These factors include the varying properties of ITO used in different testing batches, variations in enzyme and reagent concentrations due to handling issues, limited control over environmental factors such as temperature and humidity, variability between reagent batches, residual non-specific binding of proteins in plasma on the ITO, and varying enzyme activity influenced by the pH or ion concentrations of plasma samples. These factors can be addressed in future iterations through quality controls implemented in the fabrication and assembly processes. Additionally, assay and environmental factors that have a direct impact on the captured data can be added to the input of future inference models to improve the generalizability of the concentration inference model to different batches.

The potential incorporation of our platform into immunoassay technology will open up a myriad of biomedical applications, including disease diagnosis, biomarker detection, and therapeutic drug monitoring. The demonstration of electroactive enzymatic signaling between anti-HRP and TMB (Figure S9) underscores the versatility of our platform in LFA-based immunoassay techniques. Measuring electrical signals of the commonly used clinical immunoassays by our FET biosensor offers significant advantages, such as high sensitivity, enhanced accuracy through kinetic information, and data-driven analysis. Expanding on the potential benefits, the ability of FET sensors to conduct multiplexed immunoassays could be a critical advancement, enabling the simultaneous detection of multiple biomarkers within a single sample input. This capability is essential for achieving comprehensive disease profiling, providing a more nuanced understanding of an individual's health status. By facilitating the detection of a spectrum of biomarkers within a single diagnostic measurement, our platform, once fully developed, might contribute to more holistic and efficient diagnosis and monitoring of various diseases.

**CONCLUSION**

Our research showcased a seamless integration of FETs, paper-based analytical devices, and DL methodologies, effectively addressing persistent challenges associated with FET sensors, such as sample matrix effects and variations in reaction rates, while simplifying operational complexities. The inclusion of a PSM in the FET sensing zone streamlined the sensor operation into a single-step, cost-effective testing process. The synergistic interplay between FETs' kinetic data and DL methodologies was further demonstrated through quantitative diagnostics, notably in the proof-of-concept quantification of cholesterol concentration in patient plasma samples. For blinded cholesterol tests, this approach yielded a high precision (CV < 6.46%) and a decent accuracy ($r^2$ > 0.976). The integration of immunoassays into our detection platform could potentially achieve a significant advancement in medical diagnostics, promising improved healthcare outcomes.

**EXPERIMENTAL PROCEDURES**

**Sensing Solution Preparation.** A sensing solution included enzymes, a stabilizer, and buffer solutions. 300 U/mL cholesterol esterase (Toyobo, COE-311), 300 U/mL cholesterol oxidase (Toyobo, COO-321), and 300 U/mL peroxidase (Toyobo, POX-301) were dissolved in PBS, piperazine-N,N′-bis(2-ethanesulfonic acid) (PIPES), or PBS/PIPES (45/55% ratio) buffer solution, respectively. Triton X-100 (Sigma Aldrich, SLBM3869V), tween 20 (Surf's Up surfactant Kit, K40000), and 10% BSA (Thermo Scientific, WL335677) were mixed with the enzyme solution at a 0.5% (v/v) for each.

**Cartridge Fabrication.** ITO (Sigma Aldrich, 639303) cleaned with isopropanol for 20 min was utilized as the SE. ITO was further incubated with 10% BSA solution for 4 hours to achieve a blocking layer on the ITO surface. The final sensing solution described earlier was fully spread over each PSM made of an asymmetric super micron polysulfone membrane (Pall, T9EXPPA0045S00M) with a 0.45 µm average pore size and a nitrocellulose membrane with a 0.22 µm average pore size (Sartorius, 11327-41BL). Each PSM was fully dried for 20 min using nitrogen gas and stored under silica gel for 2 hours. Dried PSM was sliced to a 6 mm diameter circle for the cartridge component. The ITO was taped on an acrylic sheet substrate (1 mm-thick, 1.5 cm by 1.5 cm) using double-sided tape. Another double-sided tape (50 µm thickness) was mounted on the ITO with an opening window for PSM placement. PSM and chamber were sequentially added on the top of double-sided tape. The ITO electrode was connected to the gate of MOSFET using an alligator clip for electrical measurements. All components, including ITO, acrylic sheet and double-sided tapes, were fabricated by a laser cutter (60 W Speedy 100 $CO_2$ laser, Trotec, USA).

**Electrical Measurement System.** The ITO of the cartridge was connected to the gate of a commercial n-type metal-oxide-semiconductor field-effect transistor (MOSFET) (CD4007UB) using an alligator clip. The same MOSFET was used over all measurements consistently. A 20 µL volume of plasma was injected into the inlet of the cartridge. An Ag/AgCl reference electrode contacted the plasma, applying the $V_G$ in a range from 0 to 3 V for all measurements. All transfer curves were measured using a Keithley 4200A semiconductor analyzer with a source-drain voltage set at 50 mV, and the $V_G$ fixed in the double-sweep mode. Transfer curves of the FET were repeatedly measured for 5 min under each plasma sample. The $V_{th}$ was calculated as the $V_G$ corresponding to an $I_D$ of 1 µA in each transfer curve. Standard pH buffer solutions were used to evaluate the pH sensitivity of the ITO in Figure S1. Each solution was removed by pipetting after each measurement. For HRP response tests in Figure S9, 10 mM $H_2O_2$ in PBS was mixed with IgG-HRP (Southern Biotech, 1030-05) on the bare ITO surface, sequentially, with increasing concentrations of IgG-HRP in a range from 16 ng/mL to 50 µg/mL in PBS. Also, TMB (Thermo Scientific, 34028) was added to goat IgG-HRP solution in a range from 16 ng/mL to 10 µg/mL in PBS before testing.

**Clinical Sample Tests.** Lithium heparin plasma from leftover patient samples collected at The University of Chicago Medical Center with cholesterol concentrations ranging from 100 to 300 mg/dL were de-identified and stored at -20 °C until use. Samples were collected under a quality assurance protocol, which qualified for an institutional review board waiver and no patient identifiers were collected. Cholesterol concentrations were quantified using the Roche CHOL2 enzymatic colorimetric assay on the c701 module of the Roche Cobas 8000 analyzer system (Indianapolis, IN, USA). After thawing, the samples were stored at 2–8°C for up to seven days. Lipoprotein-free human plasma was purchased from Kalen Biomedical, LLC as control. In order to evaluate LOD (Figure 2d), 312 mg/dL clinical plasma sample was diluted by lipoprotein-free human plasma. The CV values (Figure 2e) were obtained from at least 3 testing cartridges for the same human plasma sample.

A total of 179 plasma samples were tested within 3 testing batches, including 86 plasma samples in the first batch, 57 plasma samples in the second batch, and 36 plasma samples in the third batch. In the first testing batch, 61 plasma samples were used for training, with 17 samples for validation and 8 samples for blind testing of the deep learning model. In the second batch, 42 samples were allocated for training, with 15 samples for blind testing, and in the third batch, 29 samples were reserved for training, with 7 samples for blind testing. This split was dictated by the uniform coverage of cholesterol concentration (in 100-300 mg/dL range).

**Data Processing and Deep Learning Analysis.** For each sample, the transfer curves of the FET sensor were repeatedly measured over 49 cycles with 7 sec per cycle (i.e., a total of 343 s period). Before applying DL-based analysis, the first captured cycle was subtracted from all 49 cycles within the raw heatmap (Figure S6b), yielding 48 cycles within the heatmap, termed signal heatmap (Figure 3c). For DL analysis, the signal heatmap was converted into a 1D array and input into the processing neural network. The neural network architecture was optimized through a 4-fold cross-validation on the validation set, and the optimal model represented a shallow neural network with a fully-connected architecture with 3 hidden layers (128, 64 and 32 units), each followed by batch normalization and 0.5 dropout. All three layers used ReLU activation functions and L2 regularization. The loss function (L) was MSE compiled with Adam optimizer, a learning rate of $10^{-3}$, and a batch size of 5, i.e.,

$$L = \frac{1}{N}\sum_{i=1}^{N}(y_i - y'_i)^2,$$

where $y_i$ are the ground truth analyte concentrations, $y'_i$ are the predicted concentrations, and $N$ is the batch size.

The input signals into the neural network were further optimized by selecting a subset of current values from the total operating range (i.e., 14-343 s time range and 0-3 V $V_G$ range). The optimization was done in two steps through a 4-fold cross-validation (see Deep learning-based optimization of the kinetic data for more details) on 17 samples from the validation set (see Clinical Sample Tests section for more details). This optimized model architecture (i.e., the model with optimal input subset and architecture) was further used at the blind testing phase.

The blind testing set included 30 samples (not seen during network optimization) from three different testing batches. For each batch, the final optimized models were independently trained using samples from the same batch (see Clinical Sample Tests). Training times for batches 1 to 3 were 113 s, 143 s, and 145 s, respectively. Irrespective of the batch number, blind testing of the trained model averaged 110 ms per sample for a batch size of 1, and this time decreased to 35 ms per sample when using a batch size of 10. Data preprocessing and training/testing of neural networks were performed in Python, using OpenCV and TensorFlow libraries. Training/testing of the neural networks was done on a desktop computer with a GeForce GT 1080 Ti (NVIDIA).

**Deep learning-based Optimization of the Kinetic Data.** The neural network input optimization process was performed through a 4-fold cross-validation on the validation set and was conducted in two steps: first, optimizing the $V_G$ subset within 0-3 V range (Figure 4a-c), and second, optimizing the time window within 14-343s range (Figure 4d-f). In each step, the optimal model was selected based on the MSE and $r^2$ values between the predicted and ground truth cholesterol concentrations for 17 samples from the validation dataset (see the Clinical Sample Tests subsection for further details on the split between training, validation, and testing sets).

At the first step, the optimal $V_G$ operating range was determined to be between 1.15 V and 2.45 V centered at 1.8 V (Figure 4b). The predictions generated by the model with the optimal $V_G$ window exhibited a strong correlation with the ground truth with an $r^2$ of 0.907, however a CV of 20.7% was still high (Figure 4c). To further enhance the performance, we determined the optimal time range for a fixed optimal $V_G$ window by evaluating MSE and $r^2$ maps generated on the same validation dataset with 17 samples (Figure 4d). The optimal time range based on lower MSE and higher $r^2$ was selected between 91 s and 119 s centered at 105 s, reducing the overall assay operation to < 2.5 minutes (Figure 4e). The predictions of the model with optimized $V_G$ and time subsets on the validation set showed an $r^2$ value of 0.954 and a CV of 11.4% with respect to ground truth measurements (Figure 4f), and the model with this input subset was further used during the blind testing stage.

**Author Contributions**

Electrical measurements, device fabrication, and interpretation of data were carried out by H.-J. Jang, W. Zhuang, and H. Joung. Computational analysis was carried out by A. Goncharov. The manuscript was prepared by H.-J. Jang, H. Joung., A. Goncharov, A. Ozcan, and J. H. Chen. Plasma samples were prepared and evaluated by A. Kanegusuku, C. Chan, and K.-T. J. Yeo. All authors edited the manuscript and commented on it. The project was supervised by A. Ozcan and J. H. Chen.

**Declaration of interests**

The authors have a pending patent application on the presented technology.

**Acknowledgements**

This work was financially supported by a University of Chicago Faculty Start-up fund and the US National Science Foundation (NSF) PATHS-UP Engineering Research Center (Grant #1648451).

**Figure captions**

**Figure 1.** Schematic images of (a) the diagnostic platform combining the FET detection system with an actual photo of the components of a single-use paper-based analytical cartridge. (b) Detection mechanism of cholesterol around PSM and SE within the cartridge. (c) Overview of DL signal processing framework.

**Figure 2.** (a) Representative real-time $V_{th}$ curves in response to enzyme reactions based on the cholesterol concentrations in human plasma. (b) Initial $V_{th}$ distributions of cartridges dried with different buffer components such as PBS, PIPES/PBS, PIPES. CV values of initial $V_{th}$ were compared. (c) $\Delta V_{th}$ variation of the cartridge with bare ITO and BSA/ITO for the injection of lipoprotein-free plasma. $\Delta V_{th}$ was defined as the difference between the initial $V_{th}$ and $V_{th}$ for a specific time of each cartridge. (d) LOD evaluation measured by using a diluted clinical plasma sample with lipoprotein-free plasma. (e) $\Delta V_{th}$ distribution of 178 testing cartridges with plasma samples of varying cholesterol concentrations. (f) CV values of $\Delta V_{th}$ in Figure 2(e) calculated from at least 3 repeated tests for the same plasma.

**Figure 3.** (a) Schematic of transfer curve changes over different stages of the enzymatic reaction on the cartridge. (b) Representative $G_m$ variation over the enzymatic reaction. (c) Transfer curve heatmaps after subtraction of the first cycle signal. (d) Comparison of cholesterol concentrations predicted by the neural network using the pure signal and raw heatmap.

**Figure 4**. (a) MSE and $r^2$ maps for the validation dataset from models with different $V_G$ subsets; (b) Optimal $V_G$ subset selected as a local extremum on MSE and $r^2$ maps. (c) Predictions on the validation dataset for the model with the optimal $V_G$ subset; (d) MSE and $r^2$ maps for the validation dataset from models with different time subsets. (e) Optimal time subset within $V_G$ subset selected as local extremum on MSE and $r^2$ maps. (f) Model predictions on the validation dataset for the model with optimal $V_G$ and time subsets. (g) Final model predictions on the blind testing dataset composed of 30 clinical samples from 3 different testing batches. (h) $r^2$ values expanded over 3 testing batches for models with different input subsets. (i) CV values for the optimal model expanded over different cholesterol ranges for blind tests.

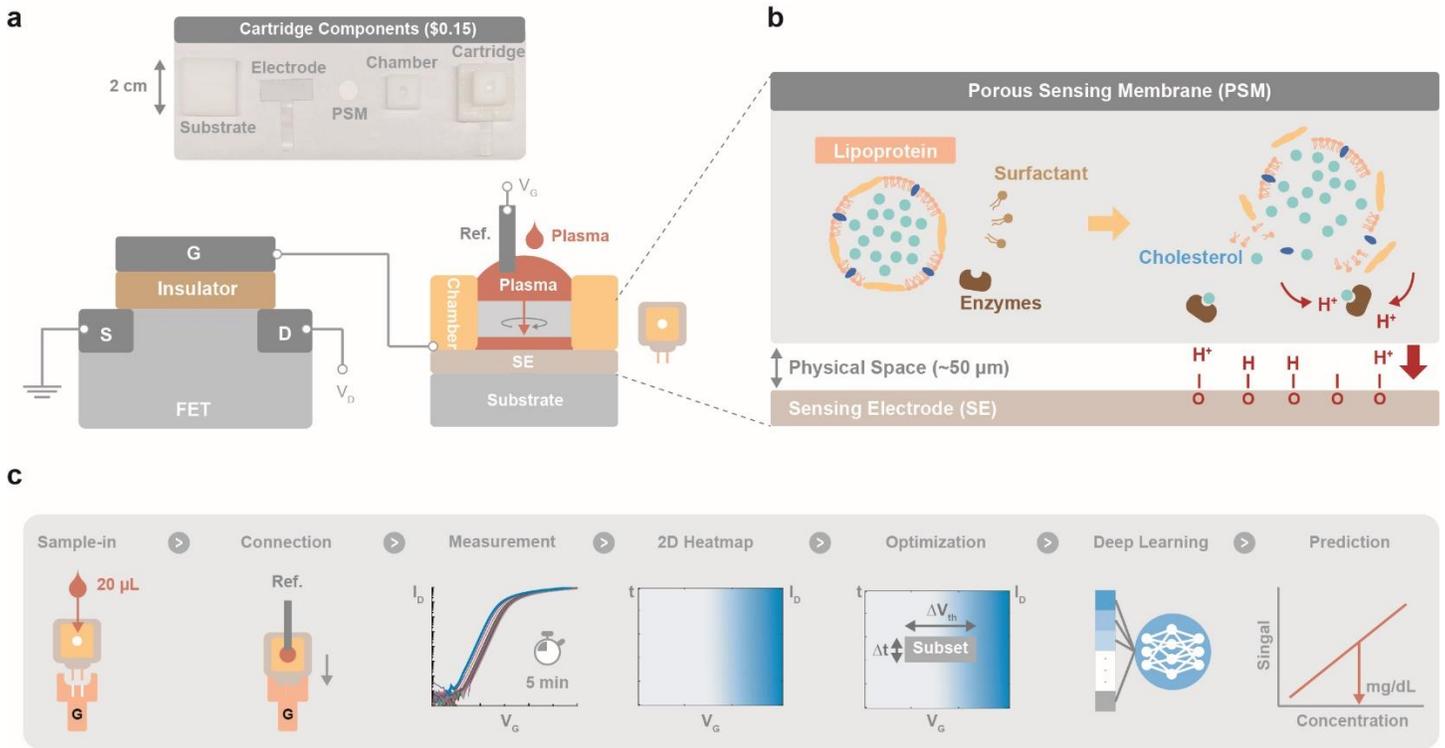

Figure 1

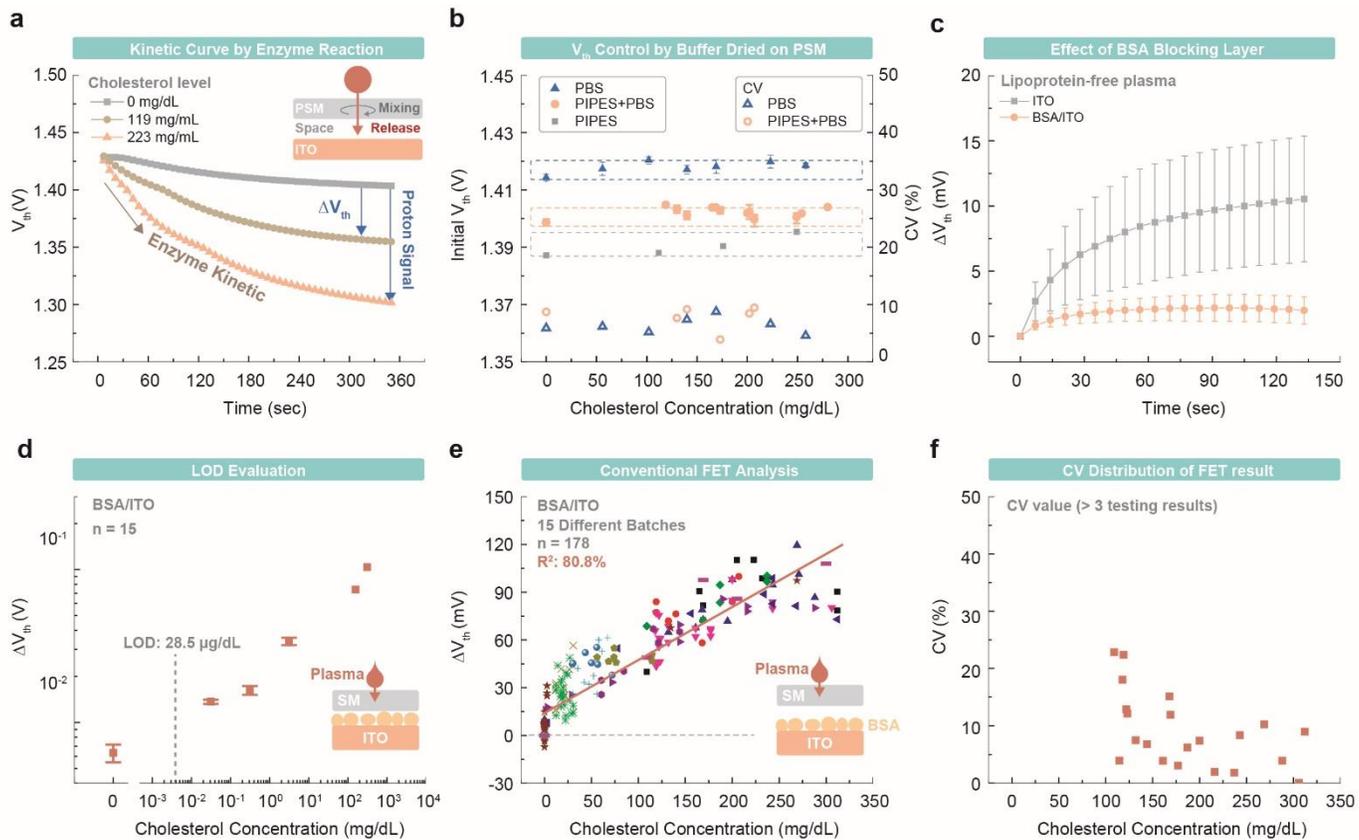

Figure 2

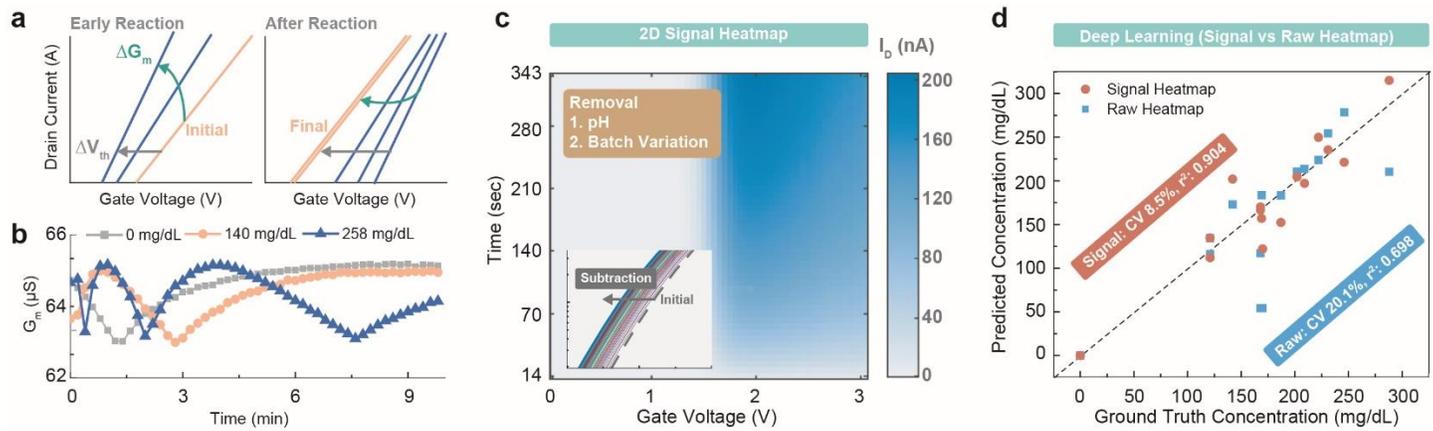

Figure 3

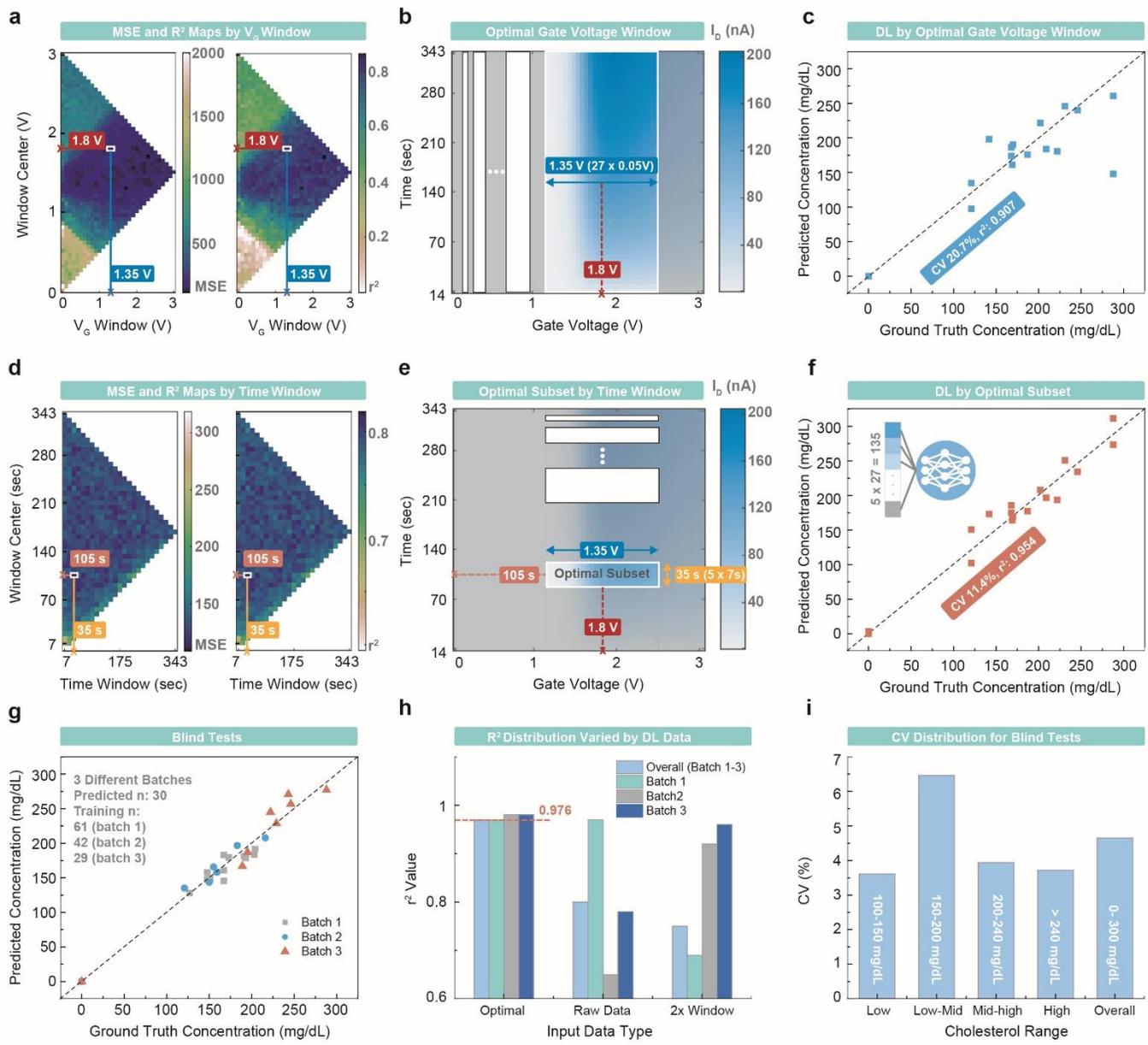

Figure 4